\documentclass[prc,aps]{revtex}
\usepackage[T1]{fontenc}
\usepackage[latin1]{inputenc}
\usepackage{graphics}
\usepackage[nofiglist,notablist]{endfloat}
\begin{document}

\draft

\title{Description of quadrupole collectivity in \protect$ N\approx 20 $ nuclei
with techniques beyond the mean field.}

\author{R.R. Rodríguez-Guzmán, J.L. Egido and L.M. Robledo}

\address{Departamento de Física Teórica C-XI, Universidad Autónoma de Madrid, 28049-Madrid,
Spain.}

\date{\today{}}

\maketitle
\begin{abstract}
Properties of the ground and several collective excited states of the light
nuclei $ ^{30,32,34}Mg $ are described in the framework of the angular momentum
projected Generator Coordinate Method using the quadrupole moment as collective
coordinate and the Gogny force as the effective interaction. The calculated
excitation energies and $ B(E2) $ transition probabilities agree reasonably
well with experiment. The results clearly indicate that both the restoration
of the rotational symmetry and the quadrupole dynamics are key ingredients for
the description of the properties of the above mentioned nuclei. 
\end{abstract}

\section{Introduction}

Nowadays, the region of neutron-rich nuclei around $ N=20 $ is the subject
of active research both in the experimental and theoretical side. The reason
is the strong experimental evidence towards the existence of quadrupole deformed
ground states in this region. The existence of deformed ground states implies
that $ N=20 $ is not a magic number for the nuclei considered, opening up
the possibility for a better understanding of the mechanisms behind the shell
structure in atomic nuclei. In addition, the extra binding energy coming from
deformation can help to extend thereby the neutron drip line in this region
far beyond what could be expected from spherical ground states. Among the variety
of available experimental data, the most convincing evidence for a deformed
ground state is found in the $ ^{32}Mg $ nucleus where both the excitation
energy of the lowest lying $ 2^{+} $ state~\cite{32Mg.E2+} and the $ B(E2,0^{+}\rightarrow 2^{+}) $
transition probability~\cite{32Mg.BE2} have been measured. Both quantities
are fairly compatible with the expectations for a rotational state. Theoretically,
from a shell model point of view, the deformed ground states are a consequence
of the lower energies of some intruder $ 2p-2h $ neutron excitations into
the \emph{fp} shell as compared to the pure \emph{sd} configuration \cite{Poves.87-94}.
In terms of the mean field picture of the nucleus, a quadrupole deformed ground
state only appears after taking into account the zero point rotational energy
correction to the mean field energy \cite{Campi.75,ray.6,ray.7,Berger.93,Rein.99}.

In a previous paper \cite{Rayner.2000} we have computed angular momentum projected
(AMP) energy landscapes, as a function of the mass quadrupole moment, for the
nuclei $ ^{30-34}Mg $ and $ ^{32-38}Si $. We have found that the projection
substantially changes the conclusions extracted from a pure mean field calculation.
In all the nuclei considered, exception made of $ ^{34}Mg, $ two coexistent
configurations (prolate and oblate) have been found with comparable energy indicating
thereby that configuration mixing of states with different quadrupole intrinsic
deformation had to be considered. The purpose of this paper is to study the
effect of such configuration mixing for the nuclei $ ^{30-34}Mg. $ The $ Si $
isotopes have been disregarded in this work as there are indications \cite{Berger.93}
that triaxiality effects could be relevant for the description of their ground
states and, for the moment, our calculations are restricted to axially symmetric
($ K=0 $) configurations. In our calculations we have used the Gogny force
\cite{Dech_Gogny.80} (with the D1S parameterization \cite{Berger.84_D1S})
which is known to provide reasonable results for many nuclear properties like
ground state deformations, moments of inertia, fission barrier parameters, etc,
all over the periodic table. As the results presented in this paper will show,
this force is also suited for the description of quadrupole collectivity in
$ N\approx 20 $ nuclei. Additional results for $ ^{32}Mg $ with older
parameterizations of the Gogny force are also discussed. Finally, let us mention
that similar calculations to the ones discussed here using the Skyrme interaction
have recently been reported \cite{Heenen.99}.

\section{Theoretical framework}

To compute the properties of the ground and several collective excited states
of the nuclei considered in this paper we have used the angular momentum projected
Generator Coordinate Method (AMP-GCM) with the mass quadrupole moment as generating
coordinate. To this end, we have used the following ansatz for the $ K=0 $
wave functions of the system \begin{equation}
\left| \Phi ^{I}_{\sigma }\right\rangle =\int dq_{20}f^{I}_{\sigma }(q_{20})\hat{P}^{I}_{00}\left| \varphi (q_{20})\right\rangle 
\end{equation}
 In this expression $ \left| \varphi (q_{20})\right\rangle  $ is the set
of axially symmetric (i.e. $ K=0 $) Hartree-Fock-Bogoliubov (HFB) wave functions
generated by constraining the mass quadrupole moment to the desired values $ q_{20}=\left\langle \varphi (q_{20})\right| z^{2}-1/2(x^{2}+y^{2})\left| \varphi (q_{20})\right\rangle  $
(please, notice that this definition is a factor $ 1/2 $ smaller than the
usual definition of the intrinsic quadrupole moment). The intrinsic wave functions
$ \left| \varphi (q_{20})\right\rangle  $ have been expanded in a Harmonic
Oscillator (HO) basis containing 10 major shells and with equal oscillator lengths
to make the basis closed under rotations \cite{Robledo.94}. The rotation operator
in the HO basis has been computed using the formulas of \cite{Nazmitdinov.96}. 

The operator 
\begin{equation}
\hat{P}^{I}_{00}=\frac{2I+1}{8\pi ^{2}}\int d\Omega d^{I}_{00}(\beta )e^{-i\alpha \hat{J}_{z}}e^{-i\beta \hat{J}_{y}}e^{-i\gamma \hat{J}_{z}}
\end{equation}
is the usual angular momentum projector with the $ K=0 $ restriction \cite{Hara.95}
and $ f^{I}_{\sigma }(q_{20}) $ are the {}``collective wave functions{}''
solution of the Hill-Wheeler (HW) equation 
\begin{equation}
\int dq_{20}^{,}{\mathcal{H}}^{I}(q_{20},q^{,}_{20})f^{I}_{\sigma }(q_{20}^{,})=E^{I}_{\sigma }\int dq_{20}^{,}{\mathcal{N}}^{I}(q_{20},q^{,}_{20})f^{I}_{\sigma }(q_{20}^{,}).
\end{equation}
 In the equation above we have introduced the projected norm $ {\mathcal{N}}^{I}(q_{20},q^{,}_{20})=\left\langle \varphi (q_{20})\right| \hat{P}^{I}_{00}\left| \varphi (q_{20}^{,})\right\rangle  $,
and the projected hamiltonian kernel $ {\mathcal{H}}^{I}(q_{20},q^{,}_{20})=\left\langle \varphi (q_{20})\right| \hat{H}\hat{P}^{I}_{00}\left| \varphi (q_{20}^{,})\right\rangle  $.
As the generating states $ \hat{P}^{I}_{00}\left| \varphi (q_{20})\right\rangle  $
are not orthogonal, the {}``collective amplitudes{}'' $ f^{I}_{\sigma }(q_{20}) $
cannot be easily interpreted. This drawback can be easily overcome by introducing
\cite{Ring_Suck.80} the so-called {}``natural{}'' states \[
\left| k^{I}\right\rangle =(n^{I}_{k})^{-1/2}\int dq_{20}u^{I}_{k}(q_{20})\hat{P}^{I}_{00}\left| \varphi (q_{20})\right\rangle \]
 which are defined in terms of the eigenstates $ u^{I}_{k}(q_{20}) $ and
eigenvalues $ n^{I}_{k} $ of the projected norm, i.e. $ \int dq^{,}_{20}{\mathcal{N}}^{I}(q_{20},q^{,}_{20})u^{I}_{k}(q^{,}_{20})=n^{I}_{k}u^{I}_{k}(q_{20}) $.
The correlated wave functions $ \left| \Phi ^{I}_{\sigma }\right\rangle  $
are written in terms of the natural states as \[
\left| \Phi ^{I}_{\sigma }\right\rangle =\sum _{k}g^{\sigma ,I}_{k}\left| k^{I}\right\rangle \]
 where the new amplitudes $ g^{\sigma ,I}_{k} $ have been introduced. In
terms of the amplitudes $ g^{\sigma ,I}_{k} $ the collective wave functions
\begin{equation}
\label{g}
g_{\sigma }^{I}(q_{20})=\sum _{k}g^{\sigma ,I}_{k}u^{I}_{k}(q_{20})
\end{equation}
 are defined. They are orthogonal and therefore their module squared has the
meaning of a probability. The introduction of the natural states also reveals
a particularity of the HW equation: if the norm has eigenvalues with zero value
they have to be removed for a proper definition of the natural states (i.e.
linearly dependent states are removed from the basis). In practical cases, in
addition to the zero value eigenvalues also the eigenvalues smaller than a given
threshold have to be removed to ensure the numerical stability of the solutions
of the HW equation. In order to account for the fact that the mean value of
the number of particles operator $ \left\langle \Phi _{\sigma }^{I}\right| \hat{N}_{\tau }\left| \Phi _{\sigma }^{I}\right\rangle  $
($ \tau =\pi ,\nu  $) usually differs from the nucleus' proton and neutron
numbers we have followed the usual recipe \cite{Hara.82,Bon.90} of replacing
the hamiltonian by $ \hat{H}-\lambda _{\pi }(\hat{N}_{\pi }-Z)-\lambda _{\nu }(\hat{N}_{\nu }-N) $
where $ \lambda _{\pi } $ and $ \lambda _{\nu } $ are chemical potentials
for protons and neutrons respectively. 

Concerning the density dependent part of the Gogny force we have used the usual
prescription already discussed in Refs. \cite{Bon.90,Rayner.2000,Valor.96}.
It amounts to use the density
\begin{equation}
\rho (\vec{r})=\frac{\left\langle \varphi (q_{20})\right| \hat{\rho }e^{-i\beta \hat{J}_{y}}\left| \varphi (q_{20}^{,})\right\rangle }{\left\langle \varphi (q_{20})\right| e^{-i\beta \hat{J}_{y}}\left| \varphi (q_{20}^{,})\right\rangle }
\end{equation}
 in the density dependent part of the interaction when the evaluation of $ \left\langle \varphi (q_{20})\right| \hat{H}e^{-i\beta \hat{J}_{y}}\left| \varphi (q_{20}^{,})\right\rangle  $
is required in the calculation of the projected hamiltonian kernels.

It has to be kept in mind that the solution of the HW equation for each value
of the angular momentum $ I $ determines not only the ground state ($ \sigma =1) $,
which corresponds to the Yrast band, but also excited states ($ \sigma =2,3,\ldots  $)
that, with the set of generating wave functions used in these calculations,
could correspond to solutions with a different deformation from the one of the
ground state and/or to quadrupole vibrational bands.

Finally, let us mention that, as the intrinsic wave functions $ \left| \varphi
(q_{20})\right\rangle  $ are determined before the projection onto angular
momentum, the procedure described above is of the {}``projection after
variation{}'' (PAV) type. It is well known \cite{Ring_Suck.80} that the PAV
method yields the wrong moments of inertia, at least in the translational case,
and a way to cure this deficiency is to consider a {}``projection before
variation{}'' (PBV) which is much more difficult to implement because the
intrinsic wave functions have to be determined for each value of the angular
momentum $ I $ using the Ritz variational principle on the projected energy
(see \cite{Schmid.87} for the application of PBV with small configuration
spaces). To illustrate the consequences of the PBV method it is convenient to
consider a strongly deformed intrinsic configuration $ \left| \varphi
(q_{20})\right\rangle  $ as in this case it is possible to obtain
\cite{Ring_Suck.80} an approximate expression for the (PAV) projected energy 
$ E_{PAV}(I)=<H>-\frac{<\vec{J}^{2}>}{2\mathcal{J}_{\mathrm{Y}}}+\frac{
\hbar^{2}I(I+1)}{2\mathcal{J}_{\mathrm{Y}}} $ where $ \mathcal{J}_{\mathrm{Y}} $
is the Yoccoz (Y) moment of inertia. In this expression we recognize the
rotational energy correction $ \frac{<\vec{J}^{2}>}{2\mathcal{J}_{\mathrm{Y}}}
$ and the usual rotor-like expression for the energy of the band $
\frac{\hbar ^{2}I(I+1)}{2\mathcal{J}_{\mathrm{Y}}} $. It was shown in
\cite{Villars.71} (see also \cite{Friedman.70}) that starting from the
projected energy and making an approximate projection before variation (PBV)
one obtain for the energy of the rotational band the following expression $
E_{PBV}(I)=<H>-\frac{<\vec{J}^{2}>}{2\mathcal{J}_{\mathrm{Y}}}+\frac{\hbar
^{2}I(I+1)}{2\mathcal{J}_{\mathrm{TV}}} $ where $ \mathcal{J}_{\mathrm{TV}}
$ is the Thouless-Valatin (TV) moment of inertia. This implies that for the
determination of the zero point rotational energy correction (which is very
important as it can dramatically change the energy landscape as a function of
the quadrupole moment) one has to use the Yoccoz moment of inertia (i.e. PAV is
good) but for the moment of inertia of the band one has to use the
Thouless-Valatin expression or carry out a full PBV calculation. 

Taking into account that, in the limit of strong deformation the PBV for the
restoration of the rotational symmetry yields to the well known Self Consistent
Cranking (SCC) method, a possible way to improve the AMP-GCM would be to
consider for the intrinsic states a set of wave functions 
$ \left| \varphi^{I}(q_{20})\right\rangle  $ solution of the SCC-HFB equations for each spin
$ I. $ However, this would lead to a triaxial projection which is extremely
time consuming and also to the issue of how to handle configurations with $
q_{20} $ values close to sphericity where the SCC-HFB is no longer a good
approximation to the PBV theory. 

In order to explore the effect of the PBV in our calculations we will restraint
ourselves to perform SCC-HFB calculations for selected configurations and compare
the results with those of an AMP calculation on those configurations in order
to extract the SCC and Yoccoz moments of inertia. The result of the comparison
is that the AMP gamma ray energies are typically a factor 1.4 bigger than the
selfconsistent ones and therefore a way to incorporate the effects of PBV would
be to quench the bands generated by the AMP-GCM by a factor $ 1/1.4\approx 0.7 $.
From a physical point of view it is rather simple to understand why the AMP
rotational band energies are higher than the SCC ones. For the sake of simplicity
we will concentrate on the $ 0^{+} $ and $ 2^{+} $ states. The effect
of the PBV on the $ 0^{+} $ state is to incorporate into the corresponding
intrinsic state admixtures of two, four, etc quasiparticle configurations coupled
to $ K=0. $ For the $ 2^{+} $ state we can also mix $ K=1 $ and $ K=2 $
multiquasiparticle configurations that make the variational space bigger and
therefore leads to a higher energy gain for the $ 2^{+} $ state as compared
to the energy gain of the $ 0^{+} $ state reducing thereby the corresponding
$ 2^{+} $ gamma ray energy.

\section{Discussion of the results}

\subsection{Mean field and angular momentum projected energies}

In figure 1 we have plotted the $ I=0\hbar ,2\hbar ,4\hbar ,6\hbar  $ and
$ I=8\hbar  $ projected energies 
$ E^{I}(q_{20})={\mathcal{H}}^{I}(q_{20},q_{20})/
{\mathcal{N}}^{I}(q_{20},q_{20}) $
as a function of $ q_{20} $ for the nuclei $ ^{30,32,34}Mg $. The HFB
energies have also been plotted for comparison. The projected energy curves
can be regarded as the potential energies felt by the quadrupole collective
motion and therefore give us indications of where the collective wave functions
will be concentrated. 

Before commenting the physical contents of the curves we have to mention that,
except for the $ I=0\hbar  $ curves, several values around $ q_{20}=0 $
are omitted. They correspond to intrinsic configurations with a very small value
of the norm $ {\mathcal{N}}^{I}(q_{20},q_{20}) $, that is, to configurations
whose $ I=2\hbar ,4\hbar ,\ldots  $ contents are very small. As a consequence,
the evaluation of the projected energies in these cases is vulnerable to strong
numerical inaccuracies. Fortunately, the smallness of their projected norms
guarantees that these configurations do not play a role in the configuration
mixing calculation (the associated norm eigenvalues $ n^{I}_{k} $ are very
small) and therefore can be safely omitted.

Coming back to the projected energy surfaces, we observe that for $ I=0\hbar  $
and $ 2\hbar  $ a prolate and an oblate minima appear with almost the same
energy for the nucleus $ ^{30}Mg $ whereas the prolate minimum becomes deeper
than the oblate one for $ ^{32,34}Mg. $ For increasing spins either the prolate
minimum becomes significantly deeper than the oblate one or the oblate minimum
is washed out. The prolate minima are located, for all nuclei and spin values,
around $ q_{20}=1b $ that corresponds to a $ \beta  $ deformation parameter
of $ 0.4. $ On the other hand, the HFB energy curves show a behavior rather
different from the $ I=0\hbar  $ projected curves showing a spherical minimum
for $ ^{30,32}Mg $ and a prolate one for $ ^{34}Mg. $

\begin{figure}
\resizebox*{0.9\textwidth}{!}{\rotatebox{-90}{\includegraphics{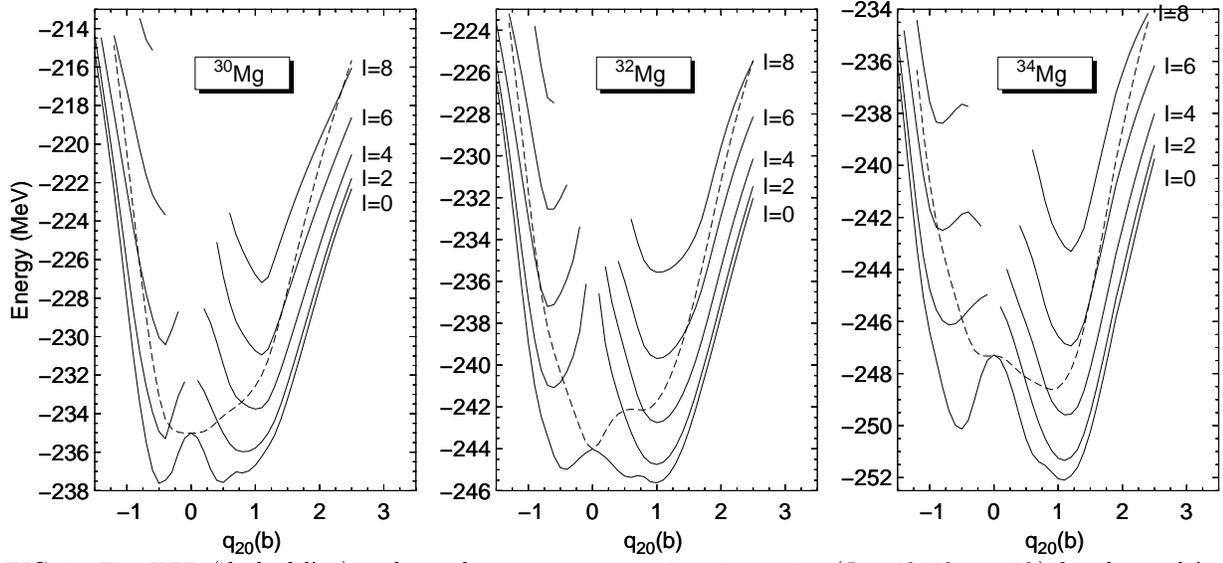}}}

\caption{The HFB (dashed line) and angular momentum projected energies \protect\protect$ (I=0\hbar ,2\hbar ,\ldots ,8\hbar )\protect $
for the nuclei considered. See text for further comments.}
\end{figure}

To disentangle the relevant configurations of the intrinsic wave functions we
have computed their spherical orbit occupancies which are given by 
\begin{equation}
\label{occ}
\nu (nlj)=\left\langle \varphi (q_{20})\right| \sum _{m}c^{+}_{nljm}c_{nljm}\left| \varphi (q_{20})\right\rangle 
\end{equation}
 where $ c_{nljm} $ are the annihilation operators corresponding to spherical
harmonic oscillator wave functions. In the nucleus $ ^{32}Mg $ the neutron
$ \nu (1f_{7/2}) $ occupancy is zero for $ q_{20}=0 $ whereas it is almost
$ 2 $ at the minimum of the projected energy (i.e. $ q_{20}=1b $). The
conclusion is clear, the zero point energy associated to the restoration of
the rotational symmetry favors the configuration in which a couple of neutrons
have been promoted from the $ sd $ shell to the $ f_{7/2} $ orbit. This
is in good agreement with the Shell Model picture of deformation in these nuclei
\cite{Poves.87-94}. 

Through exhaustive mean field studies of the nucleus $ ^{32}Mg $ with several
parameterizations of the Skyrme interaction \cite{Rein.99} it has become clear
that the occurrence of deformation in this nucleus is correlated to the 
relative
position between the $ f_{7/2} $ and $ d_{3/2} $ neutron orbitals. In
our case (D1S parameterization of the Gogny interaction) the so-called $ sd-pf $
spherical shell gap for neutrons in the nucleus $ ^{32}Mg $, which is given
by $ \Delta \epsilon _{f_{7/2}-d_{3/2}}=\epsilon _{f_{7/2}}-\epsilon _{d_{3/2}} $
(with $ \epsilon  $ being the single particle energies of the spherical configuration),
takes the value $ 5.4MeV $. This value is compatible with the results of
\cite{Rein.99} and also with the value given in \cite{Utsumo.99}. Furthermore,
the $ f_{7/2}-p_{3/2} $ spherical energy gap is only $ 1.8MeV $ and therefore
we expect strong quadrupole correlations between these two orbits. The values
for other parameterizations of the Gogny force will be discussed in the last
subsection. Finally, let us mention that the quantity $ \Delta \epsilon _{f_{7/2}-d_{3/2}} $
is not well defined for the $ ^{30}Mg $ and $ ^{34} $Mg nuclei as in these
two cases we have appreciable neutron pairing correlations and only the quasiparticle
energies are meaningful.

\subsection{Angular momentum projected Generator Coordinate calculations.}

In figure 2 the collective wave functions squared $ \left| g_{\sigma }^{I}(q_{20})\right| ^{2} $
(see Eq. \ref{g}) for the two lowest solutions $ \sigma =1 $ and $ 2 $
obtained in the AMP-GCM calculations are depicted. We also show in each panel
the projected energy for the corresponding spin. We observe that the $ 0^{+}_{1} $
ground state wave functions of the $ ^{30}Mg $ and $ ^{32}Mg $ nuclei
contain significant admixtures of the prolate and oblate configurations whereas
for $ ^{34}Mg $ the wave function is almost completely located inside the
prolate well. At higher spins, however, the ground state wave functions are
located inside the prolate well in all the nuclei studied. Concerning the first
excited states ($ \sigma =2 $) we notice that in the nucleus $ ^{34}Mg $
and for spins higher than zero the collective wave functions show a behavior
reminiscent of a $ \beta  $ vibrational band: they are located inside the
prolate wells and have a node at a $ q_{20} $ value near the point where
the ground state collective wave functions attain their maximum values. Contrary
to the case of a pure $ \beta  $ band, the collective wave functions of fig.
2 are not symmetric around the node and therefore can not be considered as harmonic
vibrations. On the other hand, the $ 0^{+}_{2} $ state of $ ^{34}Mg $
is an admixture of prolate and oblate configurations and can not be considered
as a $ \beta  $ vibrational state. The same pattern is also seen in the other
two nuclei but with slight differences: the $ \beta  $ like bands appear
at spins $ 4 $ and $ 6 $ for $ ^{32}Mg $ and $ ^{30}Mg $ respectively.

\begin{figure}
\par\centering \resizebox*{0.75\textwidth}{!}{
\includegraphics{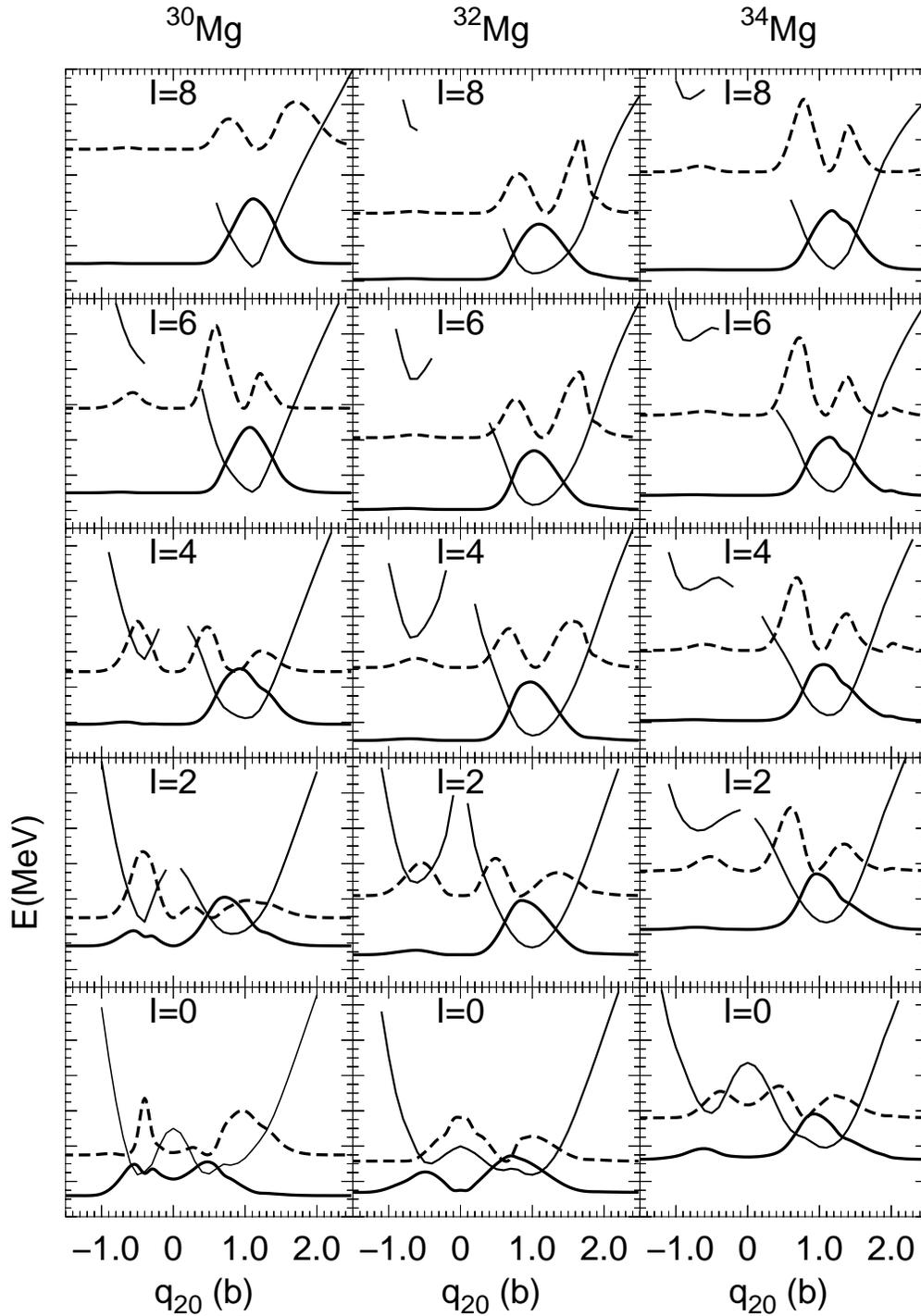} }
\caption{The collective amplitudes 
\protect$ |g^{I}_{\sigma }(q_{20})|^{2}$
(thick lines) for \protect$ \sigma =1$ (full) and 2 (dashed)
and spin values of \protect$ I=0\hbar ,\ldots ,8\hbar$
for the nuclei \protect$^{30}Mg$, \protect$^{32}Mg$
and \protect$^{34}Mg$. The projected energy curve for each
spin is also plotted (thin line). The y-axis scales are in energy units and
always span an energy interval of 13 MeV (minor ticks are 0.5 MeV apart). The
collective wave functions \protect$|g^{I}_{\sigma }(q_{20})|^{2}$
have also been plotted against the energy scale after a proper scaling 
and shifting, that is, the quantity 
\protect$ E^{I}_{\sigma }+15\times |g^{I}_{\sigma }(q_{20})|^{2}$
is the one actually plotted. With this choice of scales we can read from the
figure the energy gain due to the quadrupole fluctuations by considering the
position of the wave functions' tail relative to the projected curve.}
\end{figure}

It is also worth pointing out that from the position of the tails of the collective
wave functions relative to the projected energies (see figure caption) we can
read the energy gain due to considering the quadrupole fluctuations. The energy
gain is maximal at $ I=0\hbar  $ (0.9, 1 and 0.7 MeV for $ ^{30}Mg $,
$ ^{32}Mg $ and $ ^{34}Mg $ respectively) and quickly decreases with spin
reflecting the narrowing of the projected wells with spin. The $ S(2n) $
separation energies are now $ 7.8MeV $ and $ 6.13MeV $ for $ ^{32}Mg $
and $ ^{34}Mg $ respectively to be compared to the values obtained with the
angular momentum projection \cite{Rayner.2000} alone ($ 7.65MeV $ and $ 6.39MeV $)
and with the experimental values of $ 8.056MeV $ and $ 6.896MeV $.

\begin{table}[tb]
\centering 
\begin{tabular}{|c|c|c|c|c|c|c|c|c|c|c|c|c|} 
&
\multicolumn{4}{c|}{$ ^{30}Mg $ }&
\multicolumn{4}{c|}{$ ^{32}Mg $ }&
\multicolumn{4}{c|}{$ ^{34}Mg $ }\\
\hline 
$ I $&
 $ \left( \overline{q}_{20}\right) ^{I}_{1} $&
 $ \Sigma ^{I}_{1} $&
 $ \left( \overline{q}_{20}\right) ^{I}_{2} $&
 $ \Sigma ^{I}_{2} $&
 $ \left( \overline{q}_{20}\right) ^{I}_{1} $&
 $ \Sigma ^{I}_{1} $&
 $ \left( \overline{q}_{20}\right) ^{I}_{2} $&
 $ \Sigma ^{I}_{2} $&
 $ \left( \overline{q}_{20}\right) ^{I}_{1} $&
 $ \Sigma ^{I}_{1} $&
 $ \left( \overline{q}_{20}\right) ^{I}_{2} $&
 $ \Sigma ^{I}_{2} $\\
\hline 
0&
 0.091&
 0.558&
 0.626&
 0.685&
 0.436&
 0.692&
 0.396&
 0.601&
 0.788&
 0.691&
 0.440&
 0.723\\
\hline 
2&
 0.579&
 0.588&
 0.092&
 0.750&
 0.885&
 0.482&
 0.393&
 0.859&
 1.052&
 0.455&
 0.644&
 0.658\\
\hline 
4&
 0.962&
 0.387&
 0.215&
 0.716&
 1.012&
 0.388&
 1.041&
 0.723&
 1.136&
 0.387&
 0.819&
 0.573\\
\hline 
6&
 1.087&
 0.300&
 0.581&
 0.557&
 1.084&
 0.363&
 1.264&
 0.554&
 1.188&
 0.354&
 0.860&
 0.528\\
\hline 
8&
 1.131&
 0.289&
 1.470&
 0.562&
 1.151&
 0.368&
 1.293&
 0.515&
 1.226&
 0.332&
 0.926&
 0.560 
\end{tabular}
\caption{The average intrinsic quadrupole moment 
\protect$ \left( \overline{q}_{20}\right) ^{I}_{\sigma } $
and fluctuations \protect$ \Sigma ^{I}_{\sigma }=
\sqrt{\left( \overline{q}^{2}_{20}\right) ^{I}_{\sigma }} $
in barns for the three nuclei considered. \label{averageq2}}
\end{table}

In order to understand in a more quantitative way the collective wave functions
just discussed it is convenient to analyze the quantities 
\begin{equation}
\label{avq2}
\left( \overline{q}_{20}\right) ^{I}_{\sigma }=
\int dq_{20}\left| g_{\sigma }^{I}(q_{20})\right| ^{2}q_{20},
\end{equation}
that gives us a measure of the average deformation of the underlying intrinsic
states, and \begin{equation}
\label{dispq2}
\left( \overline{q}^{2}_{20}\right) ^{I}_{\sigma }=\int dq_{20}\left| g_{\sigma }^{I}(q_{20})\right| ^{2}q^{2}_{20}-\left( \left( \overline{q}_{20}\right) ^{I}_{\sigma }\right) ^{2}
\end{equation}
that serves as an estimation of the wave functions' spreading. The values of
$ \left( \overline{q}_{20}\right) ^{I}_{\sigma } $ and $ \Sigma ^{I}_{\sigma }=\left( \left( \overline{q}^{2}_{20}\right) ^{I}_{\sigma }\right) ^{1/2} $
corresponding to the collective wave functions of figure 2 are given in table
\ref{averageq2}. We observe that the $ 0_{1}^{+} $ and $ 2^{+}_{2} $states
of $ ^{30}Mg $ are spherical (but with strong fluctuations in the $ q_{20} $
degree of freedom) whereas the $ 2^{+}_{1} $ state is deformed ($ \beta =0.25). $
On the other hand, the $ 0^{+}_{1} $ states of $ ^{32}Mg $ and $ ^{34}Mg $
are deformed with $ \beta  $ values of $ 0.16 $ and $ 0.3 $ respectively
and have a $ \Sigma ^{I}_{1} $ value rather high, possibly due to the small
oblate hump. For spins higher than $ I=0\hbar  $ in $ ^{32,34}Mg $ and
$ I=4\hbar  $ in $ ^{30}Mg $ the ground state band is strongly deformed.
The spreading of the wave functions gets smaller for increasing spins as expected.
The excited bands also get more deformed for increasing spin, but their $ \beta  $
values never coincide with that of the ground state band. Obviously, their spreadings
are bigger than for the ground state band. 

\begin{table}[tb]
{\centering \begin{tabular}{|c||c|c|c|c||c|c|c|c|}
&
\multicolumn{4}{c||}{ $ \sigma =1 $}&
\multicolumn{4}{c|}{ $ \sigma =2 $}\\
\hline 
\hline 
&
2&
4&
6&
8&
2&
4&
6&
8\\
\hline 
\hline 
$ ^{30}Mg $&
-13.79&
-27.01&
-32.43&
-35.36&
-3.11&
-10.07&
-21.48&
-38.35\\
\hline 
$ ^{32}Mg $&
-19.15&
-26.31&
-30.09&
-31.75&
-8.63&
-23.07&
-27.22&
-29.01\\
\hline 
$ ^{34}Mg $&
-20.78&
-27.59&
-31.27&
-33.70&
-15.16&
-21.58&
-25.34&
-26.1 
\end{tabular}}
\caption{Spectroscopic quadrupole moments in 
\protect$ efm^{2}$ for \protect$ I=2\hbar ,4\hbar ,6\hbar  $
and \protect$ 8\hbar  $ and \protect$ \sigma =1$ and
2 for the three nuclei considered in this paper.\label{table_spectroscopic}}
\end{table}

A more precise definition of the quadrupole moment for protons for each of the
AMP-GCM states can be obtained from the results of the exact spectroscopic quadrupole
moments $ Q_{\sigma }(I) $ for protons (no effective charge has been used).
The values obtained for each of the wave functions $ \left| \Phi ^{I}_{\sigma }\right\rangle  $
are given in table \ref{averageq2} for the three nuclei studied and $ \sigma =1 $
and 2. All the spectroscopic moments are negative indicating prolate intrinsic
deformations. 

We can also compute the total intrinsic quadrupole moments from the spectroscopic
ones through the formula $ \left( q^{int}_{20}\right) ^{I}_{\sigma }=-\frac{2I+3}{2I}Q_{\sigma }(I)\frac{A}{Z} $
where the $ K=0 $ restriction has been taken into account and also the fact
that our $ q_{20} $ values are, by definition, a factor $ 0.5 $ smaller
than $ Q_{0} $. The factor $ A/Z $ is used to take into account the fact
that the spectroscopic quadrupole moments are given in term of the proton mass
distribution whereas the intrinsic quadrupole moments are the total ones. As
can be readily observed from table \ref{table_spectroscopic} the intrinsic
quadrupole moments obtained from the spectroscopic ones agree rather well with
the corresponding average $ \left( \overline{q}_{20}\right) ^{I}_{\sigma } $
for low spins and deviate up to a $ 20\% $ for spin $ 8\hbar  $.

\begin{table}[H]
\centering 
\begin{tabular}{|c|c|c|c|c|c|c|c|c|} 
&
\multicolumn{3}{c|}{ Calc. Energies (MeV) }&
 Exp. &
\multicolumn{3}{c|}{Calc. $ B(E2)e^{2}fm^{4} $ }&
 Exp.\\
\hline 
&
 $ 0_{1}^{+}-2_{1}^{+} $&
 $ 0_{1}^{+}-0_{2}^{+} $&
 $ 2_{1}^{+}-2_{2}^{+} $&
 $ 0_{1}^{+}-2_{1}^{+} $&
 $ 0_{1}^{+}\rightarrow 2_{1}^{+} $&
 $ 0_{1}^{+}\rightarrow 2_{2}^{+} $&
 $ 0_{2}^{+}\rightarrow 2_{2}^{+} $&
 $ 0_{1}^{+}\rightarrow 2_{1}^{+} $\\
\hline 
$ ^{30}Mg $&
 2.15&
 2.30&
 1.60&
 1.482&
 229&
 3&
 218&
 300({*})\\
\hline 
$ ^{32}Mg $&
 1.46&
 1.77&
 3.35&
 0.885&
 395&
 3.4&
 199&
 454$ \pm  $78\\
\hline 
$ ^{34}Mg $&
 1.02&
 2.35&
 3.31&
 0.75({*})&
 525&
 0&
 290&
 580({*}) 
\end{tabular}
\caption{Calculated and experimental results for excitation energies and 
\protect$ B(E2,0_{\sigma _{1}}^{+}\rightarrow 2_{\sigma _{2}}^{+}) $
transition probabilities. In the experimental data columns values marked with
an ({*}) correspond to Monte Carlo Shell Model results taken from 
Ref.~\protect\cite{Utsumo.99}.
The experimental data for the excitation energies have been taken from 
\protect\cite{32Mg.E2+}
for the \protect$ ^{32}Mg $ nucleus and from \protect\cite{30Mg.exp}
for \protect\protect$ ^{30}Mg\protect $. 
The \protect$ B(E2) $ transition probability has been taken from
\protect\cite{32Mg.BE2}.\label{table_AMP_GCM_RES}}
\end{table}
In table \ref{table_AMP_GCM_RES} the energy splittings between different states
and the $ E2 $ transition probabilities among them are compared with the
available experimental data. Concerning the $ B(E2,0_{1}^{+}\rightarrow 2_{1}^{+}) $
transition probabilities we find a very good agreement with the only known experimental
value and with the theoretical predictions of Utsumo et al. \cite{Utsumo.99}
using the Monte Carlo Shell Model (MCSM). The $ 2^{+}_{1} $ excitation energies
rather nicely follow the isotopic trend but they are larger than the experimental
values by a factor of roughly 1.5. This discrepancy could be the result of using
angular momentum projection after variation (PAV) instead of the more complete
projection before variation (PBV) that will require for each value of the angular
momentum the calculation of the generating states from the variational principle
on the projected energy (see section 2). Usually, the PBV method yields to rotational
bands with moments of inertia larger than the PAV ones \cite{Hara.82,Schmid.87}.

A full PBV is, unfortunately, extremely costly to implement with large configuration
spaces. Therefore, to estimate the effect of PBV in our results, we have resorted
to the selfconsistent cranking method which is an approximation to PBV in the
limit of large deformations. We have chosen the intrinsic state with $ q_{20}=1b $
as the most representative configuration (it approximately corresponds to the
prolate minima in all the nuclei considered) and computed the projected energies.
In addition, selfconsistent cranking calculations with the constraints $ q_{20}=1b $
in the quadrupole moment and $ \left\langle J_{x}\right\rangle =\sqrt{I(I+1)} $
in the angular momentum have been performed. The cranking results for the excitation
energies of the $ 2^{+} $ state are $ 0.548, $ 0.591 and 0.571 $ MeV $
for $ ^{34}Mg $, $ ^{32}Mg $ and $ ^{30}Mg $ respectively whereas the
corresponding projected quantities are $ 0.753 $, $ 0.873 $ and $ 0.895 $
MeV. The cranking excitation energies of the $ 2^{+} $ state are a factor
0.7 smaller than the projected ones and therefore, the effect of PBV is to increase
the moment of inertia as compared to the PAV method. If we consider the reduction
factor as significative (the $ q_{20} $ value chosen roughly corresponds
to the position of the maxima of the collective wave functions) and apply it
to our GCM results for the $ 0^{+}_{1}-2^{+}_{1} $ energy differences we
obtain the values 0.71, 1.02 and 1.50 MeV for $ ^{34}Mg $, $ ^{32}Mg $
and $ ^{30}Mg $ respectively. The new energy differences are in much better
agreement with the experimental values and the MCSM results than the uncorrected
ones. Also the corrected energy obtained for the $ 4_{1}^{+} $ state of $ ^{32}Mg $
is in good agreement with the excitation energy of $ 2.3 $ MeV of a state
of this nucleus which is a firm candidate to be the $ 4^{+} $ state belonging
to the Yrast {}``rotational band{}'' \cite{Azaiez.98}. 

Although the previous estimation can be criticized in many ways we think it
may serve as an indication that a full PBV will improve the results obtained
here. Concerning the $ B(E2) $ transition probabilities, the main effect
of the PBV will be to shift down the $ I=2\hbar ,\ldots  $ projected energy
curves keeping its shape mostly unaffected. Therefore, we do not expect big
changes both in the collective wave functions $ g^{I}_{\sigma }(q_{20}) $
and in the $ B(E) $ transition probabilities that depend on them.

\begin{figure}
\centering 
\resizebox*{0.9\textwidth}{!}
{ \rotatebox{-90}{\includegraphics{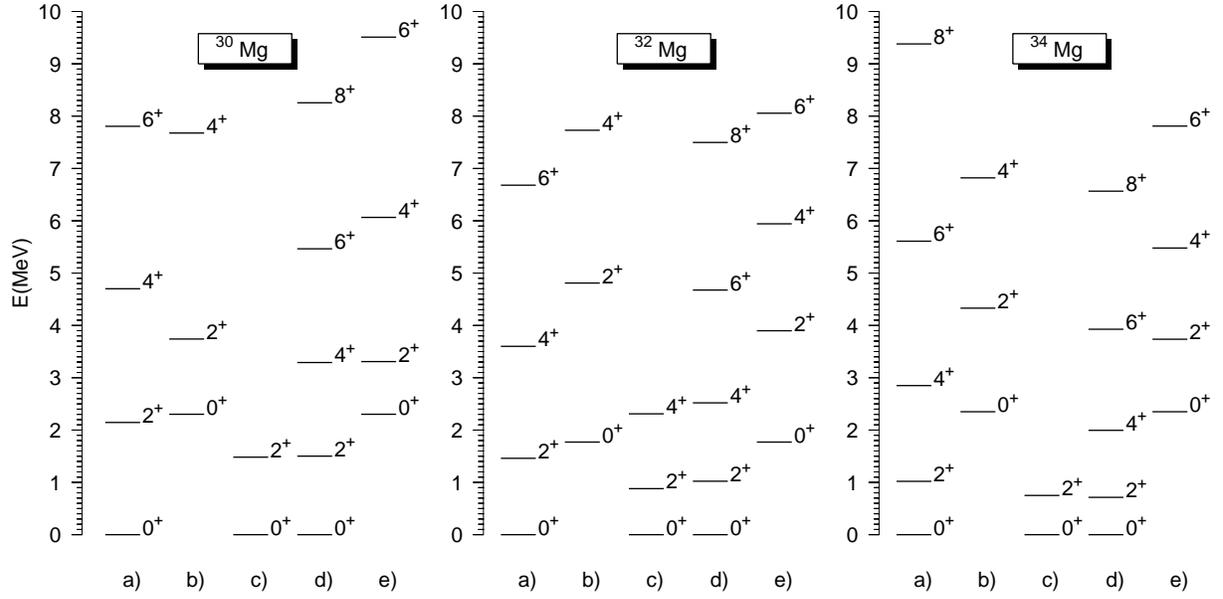} }}
\caption{Collective bands for the three nuclei studied. 
Bands a) and b) correspond to
the AMP-GCM results for the ground and first excited band. c) is the 
experimental
band (in the case of \protect$^{34}Mg$ the MCSM prediction
of \protect\cite{Utsumo.99} has been used). Finally, bands d) and e) are the
AMP-GCM results quenched by the factor 0.7 discussed in the text}
\end{figure}

Finally, the band energy diagrams for the three nuclei considered are shown
in figure 3 for states with excitation energies smaller than 10 MeV. For each
nuclei, the bands labeled (a) and (b) correspond to the AMP-GCM result for the
Yrast and excited bands, the band labeled (c) accounts for the experimental
data in $ ^{30}Mg $ and $ ^{32}Mg $ and for the MCSM result in $ ^{34}Mg $
and finally, bands (d) and (e) stand for the GCM bands quenched by the factor
0.7 previously discussed.

\subsection{Results for other parameterizations of the Gogny force.}

The occurrence of quadrupole deformation in atomic nuclei is the result of the
competition between two effects; namely, the surface energy which prevents deformation
and the quantal shell effects which, depending on the nucleus, favor quadrupole
deformation. It is therefore highly interesting to analyze the effect of these
two aspects in the results we have obtained for the nucleus $ ^{32}Mg $.
To this end we have carried out calculations with two old parameterizations
of the Gogny force; namely, the D1 and D1' parameterizations \cite{Dech_Gogny.80}.
The D1 parameterization was the one originally proposed by Gogny and the only
difference with D1' is the spin-orbit strength which is smaller for D1. As a
result one can expect that D1 will lead to a higher $ \Delta \epsilon _{f_{7/2}-d_{3/2}}=\epsilon _{f_{7/2}}-\epsilon _{d_{3/2}} $
energy gap than D1' as it turns out to be the case: the value of $ \Delta \epsilon _{f_{7/2}-d_{3/2}} $
is $ 6.37MeV $ for D1 and $ 5.37MeV $ for D1'. On the contrary, the value
of $ \Delta \epsilon _{f_{7/2}-p_{3/2}} $for D1 gets reduced from the $ 1.91MeV $
we obtain for D1' to the value $ 1.56MeV $. On the other hand, D1S has the
same spin-orbit strength as D1' (the values of $ \Delta \epsilon _{f_{7/2}-d_{3/2}} $
and $ \Delta \epsilon _{f_{7/2}-p_{3/2}} $ given in the previous paragraph
for D1' are very close to those of D1S given in a previous subsection) but its
surface energy coefficient is smaller than in D1'. The need for a reduction
of the surface energy coefficient in D1' was evident when the fission barriers
for $ ^{240}Pu $ \cite{Berger.84_D1S} were computed with the Gogny force:
they came out too high and the new D1S parameterization was proposed to cure
this deficiency of the former D1' parameterization. 

In figure 4 we have plotted the HFB energy curves (left panel) and the AMP energies
for $ I=0 $ (middle panel) and $ I=2 $ (right panel) for the three parameterizations
of the Gogny force just mentioned. We observe that the results obtained for
D1S and D1' are, apart from the overall $ 4MeV $ shift, very similar. This
similarity is a clear indication that the value of the surface energy parameter
has no influence on the results. The HFB result for D1 shows a shoulder at $ q_{20}=1b $
which is located much higher in energy than the corresponding shoulder for D1S
and D1'. As a consequence, the $ I=0 $ projected energy curve obtained with
D1 shows a very shallow minimum at $ q_{20}=0.5b $. However, the $ I=2 $
projected energy curves are very similar for the three parameterizations. The
differences found between the D1 results and the ones with the two other parameterizations
clearly indicate the sensitivity of the quadrupole properties of $ ^{32}Mg $
to the relative position of the orbits involved.

\begin{figure}
{\par\centering \resizebox*{0.8\textwidth}{!}{\includegraphics{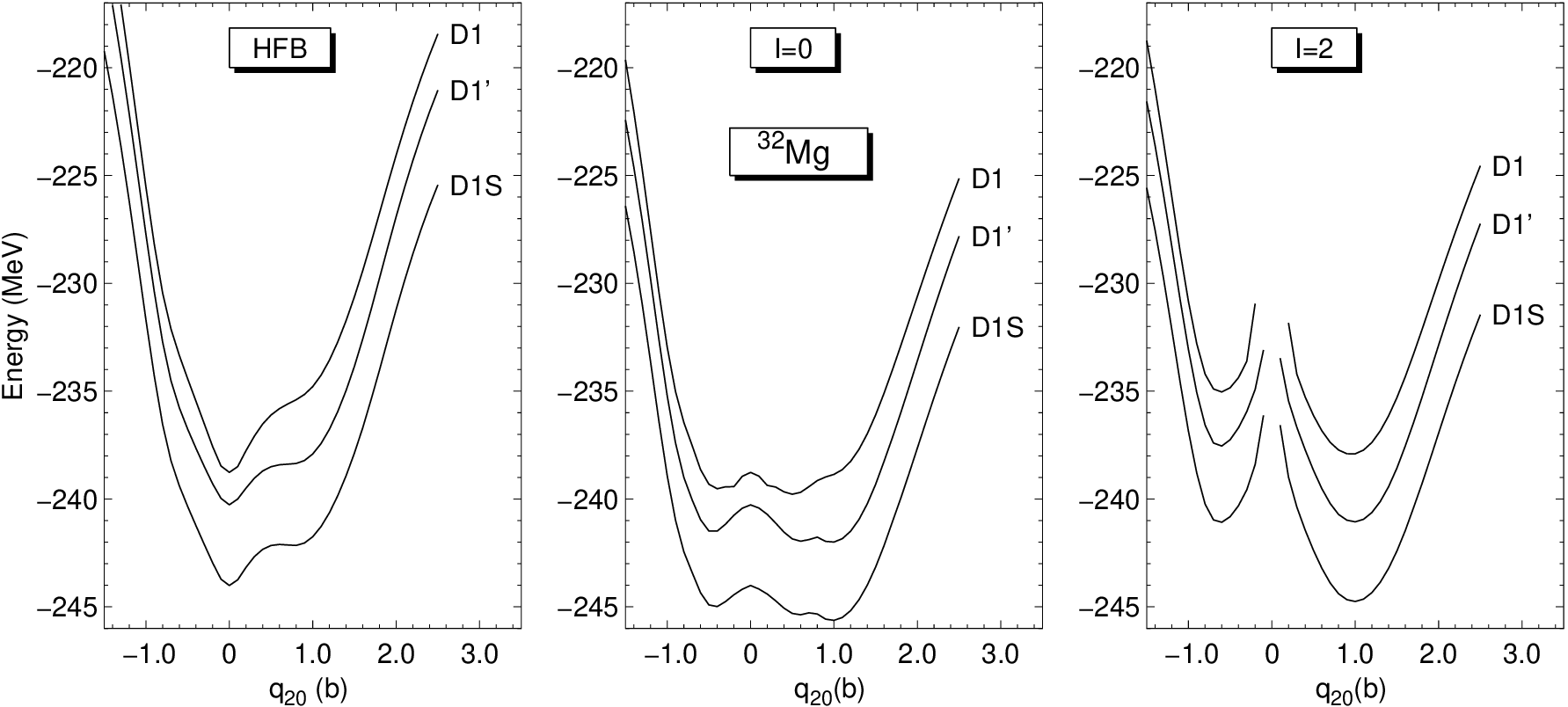}}\par}
\caption{HFB (right panel) and angular momentum projected energy curves 
(\protect$ I=0 \hbar$
middle panel, \protect$ I=2 \hbar$ left panel) as a function
of the mass quadrupole moment for the nucleus 
\protect$^{32}Mg$
and the three parameterizations of the Gogny force considered.}
\end{figure}

Finally, we have carried out the AMP-GCM calculation for the D1 and D1' parameterizations
of the force and the most important quantities obtained are summarized in table
\ref{table_parameterizations}. As expected from the projected energy curves
of figure 4 we obtain a rather small average quadrupole moment $ \left( \overline{q}_{20}\right) ^{I}_{\sigma } $
for $ \sigma =1 $ and $ I=0\hbar  $ with the D1 parameterization and bigger
ones for the two other parameterizations. However, the $ \left( \overline{q}_{20}\right) ^{I}_{\sigma } $
for $ \sigma =1 $ and $ I=2\hbar  $ are rather similar in the three cases.
The smaller value of $ \left( \overline{q}_{20}\right) ^{0}_{1} $ for the
D1 parameters gets reflected in a much smaller $ B(E2) $ transition probabilities
than for the two other parameterizations. Finally, the excitation energy of
the $ 2^{+}_{1} $ state with respect to the ground state turns out to be
significantly bigger for D1 than for the other parameter sets, being the results
of D1' and D1S in reasonable agreement. The final conclusion of this comparison
is that the energy gap $ \Delta \epsilon _{f_{7/2}-d_{3/2}} $ seems to be
a relevant parameter in order to reproduce the properties of $ ^{32}Mg $.

\begin{table}[tb]
{\centering \begin{tabular}{|c||c|c|c|c|c|c|}
&
$ \left( \overline{q}_{20}\right) ^{0}_{1} $&
$ \left( \overline{q}_{20}\right) ^{2}_{1} $&
$ B(E2,0^{+}_{1}\rightarrow 2^{+}_{1}) $&
$ E_{0^{+}_{1}-2^{+}_{1}} $&
$ \Delta \epsilon _{f_{7/2}-d_{3/2}} $&
$ \Delta \epsilon _{f_{7/2}-p_{3/2}} $\\
\hline 
D1&
0.185&
0.785&
138&
2.25&
6.37&
1.56\\
\hline 
D1'&
0.381&
0.869&
299&
1.67&
5.37&
1.91\\
\hline 
D1S&
0.436&
0.885&
385&
1.46&
5.37&
1.80
\end{tabular}}
\caption{Results of the AMP-GCM calculations for 
\protect$ ^{32}Mg $
and the parameterizations D1, D1' and D1S of the Gogny interaction. 
The average
quadrupole moments 
\protect$ \left( \overline{q}_{20}\right) ^{I}_{\sigma }$
for the ground state band and spins \protect$ 0 $ and \protect$ 2 $
are given, in barns, in the first two columns. 
In the third column the \protect$ B(E2) $
transition probabilities in \protect$ e^{2}fm^{4}$ are given.
In the fourth column the excitation energy of the 
\protect$ 2^{+}_{1} $
state with respect to the ground state is given in 
\protect$ MeV $.
Finally, in the last two columns the energy gaps 
\protect$ \Delta \epsilon _{f_{7/2}-d_{3/2}} $
and \protect$ \Delta \epsilon _{f_{7/2}-p_{3/2}} $ are given
in \protect$ MeV $.\label{table_parameterizations}}
\end{table}

\section{Conclusions}

In conclusion, we have performed angular momentum projected Generator Coordinate
Method calculations with the Gogny interaction D1S and the mass quadrupole moment
as generating coordinate in order to describe rotational like states in the
nuclei $ ^{30}Mg $, $ ^{32}Mg $ and $ ^{34}Mg $. We obtain a very well
deformed ground state in $ ^{34}Mg $, a fairly deformed ground state in $ ^{32}Mg $
and a spherical ground state in $ ^{30}Mg $. In the three nuclei, states
with spins higher or equal $ I=4\hbar  $ are deformed. The intraband $ B(E2) $
transition probabilities agree well with the available experimental data and
results from shell model like calculations. The $ 2^{+} $ excitation energies
follow the isotopic trend but come out a factor 1.5 too high as compared with
the experiment. We attribute the discrepancy to the well known deficiency of
Projection After Variation calculations of providing small moments of inertia.
However, we consider the agreement with experiment to be remarkable taking into
account that the same force used in this calculation is also able to give reasonable
values for such different quantities as fission barrier heights, moments of
inertia of superdeformed bands, the energy of octupole vibrations, etc in heavy
nuclei. The sensitivity of the results to other parameterization of the Gogny
interaction is also analyzed and the conclusion is that the D1 parameter set
fails to reproduce the properties of $ ^{32}Mg $ being the spin-orbit strength
the responsible for such failure.

\acknowledgements

One of us (R. R.-G.) kindly acknowledges the financial support received from
the Spanish Instituto de Cooperacion Iberoamericana (ICI). This work has been
supported in part by the DGICyT (Spain) under project PB97/0023.

\end{document}